\documentstyle[12pt,bezier]{article}

\def\bar{\begin{array}{rcl}}
\def\ear{\end{array}}
\def\ra{\rangle}
\def\la{\langle}
\def\beq{\begin{equation}}
\def\eeq{\end{equation}}
\def\ot{\otimes}
\def\id{\mbox{id}}
\def\ie{\mbox{\it i.e.\/ }}

\newcommand{\tr}{\triangleright}
\def\cross{\mbox{$\times \!\rule{0.3pt}{1.1ex}\,$}}
\def\smash{{\A \cross \U}}
\def\R{\mbox{$\cal R$}}

\def\A{\mbox{$\cal A$}}
\def\U{\mbox{$\cal U$}}
\newcommand{\DA}{\Delta _{\cal A}}
\newcommand{\AD}{{}_{\cal A}\Delta }
\newcommand{\UD}{{}_{\cal U}\Delta }
\newcommand{\DU}{\Delta_{\cal U}}

\def\z{\hspace*{9mm}}
\def\x{\hspace{3mm}}
\newcommand{\ad}{\stackrel{\mbox{\scriptsize ad}}{\triangleright}}
\def\Ileft{{\cal I}_{\rm left}}
\def\Gleft{{\Gamma}_{\rm left}}
\def\I{\mbox{\boldmath $i$}}
\def\Ix#1{\mbox{\boldmath $i$}_{\chi_#1}}
\def\Li{\hbox{\large\it \pounds}}
\def\Lix#1{\hbox{\large\it \pounds}_{\chi_#1}}

\def\dl{\mbox{\bf d}}

\newcommand{\om}{\mbox{$\omega$}}
\newcommand{\al}{\alpha}
\newcommand{\ep}{\epsilon}

\begin{document}
\begin{center}
August 1994	   \hfill	LMU-TPW 94-8\\

\vskip.6in

{\Large \bf Cartan Calculus: Differential Geometry for Quantum Groups}

\vskip.4in

Peter Schupp\footnote{e-mail: schupp@lswes8.ls-wess.physik.uni-muenchen.de}

\vskip.25in

{\em Sektion Physik der Ludwig-Maximilians-Universit\"at M\"unchen\\
Theoretische Physik --- Lehrstuhl Professor Wess\\
Theresienstra\ss e 37, D-80333 M\"unchen\\
Federal Republic of Germany}

\end{center}

\vskip.6in

\section{Introduction}

The topic of this lecture is differential geometry on quantum groups.
Several lecturers at this conference have talked about this subject.
For a review and a fairly extensive list of references I would like
to point the reader to the contributions of B.\ Jurco and S.\ L.\
Woronowicz in this proceedings.
Here we shall propose a new rigid framework for the so-called
 Cartan calculus
of Lie derivatives, inner derivations, functions, and forms.
The construction employs a semi-direct product of two
graded Hopf algebras, the respective super-extensions of
the deformed universal enveloping algebra and the
algebra of functions on a quantum group.
All additional relations in the
Cartan calculus follow as consistency conditions.
The approach is not based on the Leibniz rule for the
exterior derivative  and might hence also
be of interest in the recent work on its deformations.
However, given a $\dl$ that satisfies the Leibniz rule, the Cartan
identity (\ref{cartid}) follows as a theorem.

Rigorous proofs for many statements in this presentation can
be found in some form or other in \cite{Wor}. For a nice review
see e.g.\ \cite{AscCas}. For Quantum Groups and (quasitriangular)
Hopf algebras one could consult \cite{Dri,FRT}.

In the next section we would like to motivate the semi-direct
product construction by some geometrical considerations.

\subsection{Classical Left Invariant Vector Fields}

Lets recall the left-invariant classical case:
The Lie algebra is spanned by left-invariant vector fields on the group
manifold of a Lie group $G$.
These are uniquely determined by the tangent space at $1$ (the
identity of $G$). Let $g,h  \in G$.
Curves on $G$ can be naturally transported by left
(or right) translation \ie $h \mapsto g h$ ($h \mapsto h g$).
This defines  a left transport $L_{g^{-1}}$ of the tangent vectors:\x
$L_{g^{-1}}(\chi _{1}) = \tilde{\chi }_{g}$. $\chi _{1}$ is the vector field
$\chi $ at the identity
of the group and $\tilde{\chi }_{g}$ is the new vector field $\tilde{\chi }$
evaluated at the point of the group manifold corresponding to the group
element $g$; if $\chi $ is left invariant then $\chi  = \tilde{\chi }$ and
in particular
$ L_{g^{-1}}(\chi _{1}) = \tilde{\chi }_{g} = \chi _{g}.  $
An inner product for a vector field $\chi $ with a function $f$ can be defined
by acting with the vector field on the function and evaluating the
resulting function at the identity of the group:
$$
\la \chi  , f\ra := \left. \chi _{1} \tr f \right|_{1} \in k.
$$
If we know these values for all functions, we can reconstruct
the action of $\chi $ on a function $f$, $\chi _{g} \tr f|_{g}$,
at any (connected) point of the
group manifold. The construction goes as follows (see figure):\hfill\\
\mbox{}\hfill\\ \hfill
\unitlength=1.00mm
%\special{em:linewidth 0.4pt}
\linethickness{0.4pt}
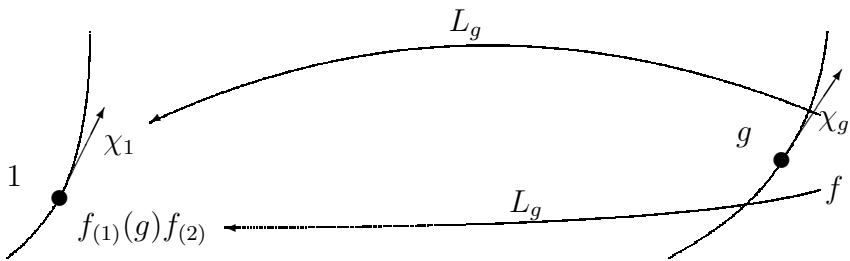
\begin{picture}(114.00,39.00)
\put(10.00,9.00){\vector(1,2){6.00}}
\put(10.00,9.00){\circle*{2.00}}
\put(106.00,14.00){\circle*{2.00}}
\put(4.00,12.00){\makebox(0,0)[cc]{1}}
\put(101.00,17.00){\makebox(0,0)[cc]{$g$}}
\put(113.00,19.00){\makebox(0,0)[cc]{$\chi_g$}}
\put(18.00,16.00){\makebox(0,0)[cc]{$\chi_1$}}
\put(113.00,10.00){\makebox(0,0)[cc]{$f$}}
\put(21.00,5.00){\makebox(0,0)[cc]{$f_{(1)}(g)f_{(2)}$}}
\put(106.00,14.00){\vector(2,3){8.00}}
\bezier{92}(10.00,9.00)(14.00,17.00)(14.00,31.00)
\bezier{44}(10.00,9.00)(7.00,4.00)(3.00,1.00)
\bezier{76}(106.00,14.00)(111.00,21.00)(112.00,31.00)
\bezier{84}(106.00,14.00)(102.00,7.00)(91.00,1.00)
\put(24.00,20.00){\vector(-2,-1){2.00}}
\bezier{388}(22.00,19.00)(65.00,39.00)(111.00,20.00)
\put(34.00,5.00){\vector(-1,0){2.00}}
\bezier{320}(32.00,5.00)(93.00,5.00)(111.00,10.00)
\put(64.00,32.00){\makebox(0,0)[cc]{$L_g$}}
\put(72.00,8.00){\makebox(0,0)[cc]{$L_g$}}
\end{picture}
\hfill \\ \mbox{} \hfill\\
We start at the point $g$, transport $f$ and $\chi $ back to the identity
by left translation and then evaluate them on each other. The result,
being a number, is invariant under translations and hence gives
the desired quantity. The left translation $Lg(f)$ of a function,
implicitly defined through $L_{g}(f)(h) = f(g h)$, finds an explicit
expression in Hopf algebra language (here and in what follows we use the
formal notation $\Delta f \equiv f_{(1)} \otimes f_{(2)}$ for the
coproduct)
$$
L_{g}(f) = f_{(1)}(g) f_{(2)},
$$
that we now use to express
$$
\begin{array}{rcl}
\left. \chi _{g} \tr f\right|_{g}
	& = & \left. L_{g}(\chi )_{1} \tr f_{(1)}(g) f_{(2)}\right|_{1}\\
        & = & \left. \chi _{1} \tr f_{(1)}(g) f_{(2)}\right|_{1}\\
	& = & f_{(1)}(g) \la \chi  , f_{(2)}\ra ,
\end{array}
$$
for a left-invariant vector field $\chi $.
If the drop $g$, we obtain the familiar expression for the action of a vector
field on a function valid on the whole group manifold
\beq
\chi  \tr f = f_{(1)}\la \chi  , f_{(2)}\ra .
\eeq
The left and right
vacua find the following geometric interpretation:
\begin{quote}
left vacuum $\la $: ``Evaluate at the identity (of the group).''\\
right vacuum $\ra $: ``Evaluate on the unit function.''
\end{quote}

\subsection{``Quantum Geometry''}

Group elements ($g$) ``an sich'' do not exist for quantum groups,
everything has to
be formulated in terms of a Hopf algebra of functions. The group
operation is replaced by the coproduct of functions.  If we take care only to
speak about functions in \A\ and its dual Hopf algebra \U, we can, however,
still develop a	geometric picture for vector fields on quantum groups.
``Points'' will be {\em labeled} by elements of $\widehat{\U}$, which is
the same as \U\ but has the opposite multiplication;
if elements of $\U$ are left-invariant then elements of
$\widehat{\U}$ are {\em right}-invariant. Lie derivatives along elements
of $\widehat{\U}$ take the place of left translations, while Lie
derivatives along elements of \U\ correspond to right translations.
Here is the quantum picture of the classical construction given
above:\\ \hfill
\unitlength=1.00mm
%\special{em:linewidth 0.4pt}
\linethickness{0.4pt}
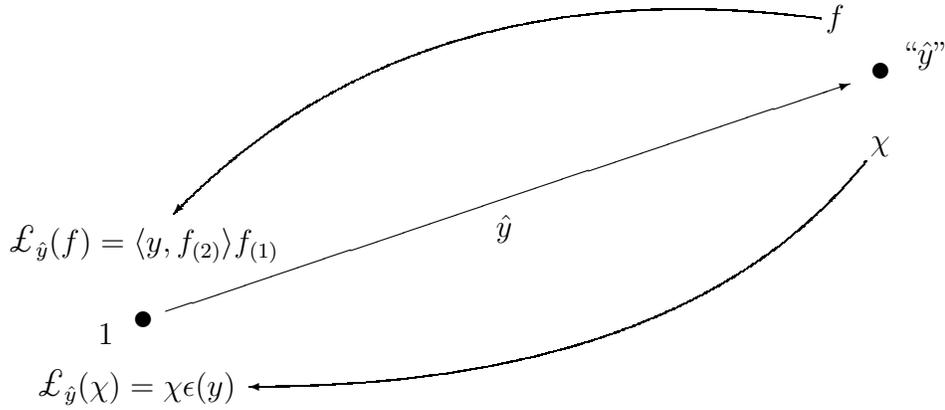
\begin{picture}(125.00,62.00)
\put(24.00,16.00){\vector(3,1){91.00}}
\put(119.00,48.00){\circle*{2.00}}
\put(21.00,15.00){\circle*{2.00}}
\put(125.00,50.00){\makebox(0,0)[cc]{``$\hat{y}$''}}
\put(16.00,13.00){\makebox(0,0)[cc]{1}}
\put(119.00,38.00){\makebox(0,0)[cc]{$\chi$}}
\put(113.00,55.00){\makebox(0,0)[cc]{$f$}}
\put(20.00,6.00){\makebox(0,0)[cc]{$\Li_{\hat{y}}(\chi) = \chi \epsilon(y)$}}
\put(21.00,25.00){\makebox(0,0)[cc]{$\Li_{\hat{y}}(f)
= \la y,f_{(2)}\ra f_{(1)}$}}
\put(69.00,27.00){\makebox(0,0)[cc]{$\hat{y}$}}
\put(36.00,6.00){\vector(-1,0){1.00}}
\bezier{384}(35.00,6.00)(93.00,6.00)(117.00,36.00)
\put(26.00,30.00){\vector(-1,-1){1.00}}
\bezier{400}(25.00,29.00)(57.00,62.00)(111.00,55.00)
\end{picture}\hfill
\\
$\Li_{\hat{y}}(x) = x \epsilon (\hat{y})$ because $x$ is
left-invariant.
Note that multiple occurrences of the same Hopf algebra element in
a single term
are unnatural. Instead one should use the parts of the coproduct of this
element.
We now compute $x \tr f$ in complete analogy to the classical case
$$
\begin{array}{rcl}
\left. x \tr f\right|_{``\hat{y}"} &=& \left. \Li_{\hat{y}_{(1)}}(x) \tr
\Li_{\hat{y}_{(2)}}(f)\right|_{1}\z(\equiv \left. \Li_{\hat{y}}(x \tr
f)\right|_{1})\\
        &=& \left. \epsilon (y_{(1)}) x
            \tr \Li_{\hat{y}_{(2)}}(f)\right|_{1}\\
        &=& \left. x \tr \Li_{\hat{y}}(f)\right|_{1}\\
	&=& \left. x \tr \la y,f_{(1)}\ra f_{(2)}\right|_{1}\\
	&=& \la y,f_{(1)}\ra\la x,f_{(2)}\ra
\end{array}
$$
or, for arbitrary $\hat{y}$:
\beq
x \tr f = f_{(1)}\la x,f_{(2)}\ra ,
\eeq
giving a geometric justification for the familiar left action of \U\
on \A.

There is also a geometric picture for the adjoint action in \U, which
can be interpreted as a quantum Lie bracket.
Recall the classical construction: Functions and hence curves on
a group manifold can be transported along a vector field. With
the curves we implicitly also transport their tangent vectors.
This transport is called the Lie derivative of a (tangent) vector
along a vector field. Classically we find it to be equal to
the commutator (Lie bracket) of the two vector fields.
In the quantum case we have
\beq
\Li_{y}(x) = y_{(1)} x S(y_{(2)}) = y \ad x.
\eeq

\subsection{Action of General Vector Fields}

Our derivation of the action of a vector field on a function in the
previous section relied on the use of left translations in conjunction
with left-invariant vector fields. For completeness we will now
consider the action of a general vector field ---
neither necessarily left or right
invariant --- on a function using alternatively left or right
translations.

Left and right coactions $\AD$, $\DA$ contain the information about
transformation properties of vector fields. Here is how a
vector field transforms (classically) if we {\em left}-transport it
from a point $g$ on the group manifold back to the identity
$$
\chi |_{g} \mapsto \chi ^{(1)'}(g)\cdot \chi ^{(2)} |_{1},\z\AD(\chi ) \equiv
\chi ^{(1)'}\ot \chi ^{(2)};
$$
here is the behavior under a {\em right} translation:
$$
\chi |_{g} \mapsto \chi ^{(1)}\cdot \chi ^{(2)'}(g) |_{1},\z\DA(\chi ) \equiv
\chi ^{(1)} \ot \chi ^{(2)'}.
$$
If we now redo the construction of the previous section for
general vector fields $\chi $, both for left and right translations, we
get the following two equivalent results for actions on functions:
\beq
\chi (f) = \underbrace{\la \chi ^{(1)},f_{(1)}\ra
\chi ^{(2)'}f_{(2)}}_{\mbox{from right translation}} = \underbrace{\chi ^{(1)'}
f_{(1)}
\la \chi ^{(2)},f_{(2)}\ra }_{\mbox{from left translation}}.
\eeq
Technically there is an ordering ambiguity for $f$ and the primed
parts of $\chi $, but this can be easily resolved by requiring
$a(f) = a f$ for $a \in \A$ in both cases; both expressions are
written as left actions. Luckily we do not have to work with these
complicated formulas --- it is sufficient to consider left-invariant
vector fields as we will see.

\subsection{Algebra of Differential Operators}

In what follows we will fix  the {\em convention}
that elements in $\U \cong  \A^{*}$ are left-invariant.
Vector fields with different behavior under transformations
can be realized as left-invariant vector fields with functional
coefficients, they hence live in $\A \ot \U$.
The left action of $x \in \U$ on products in \A\ , say $b f$, is given
via the coproduct in \U\ ,
\beq
x \tr b f  =  (b f)_{(1)} \la x,(b f)_{(2)}\ra
 =  b_{(1)} \la x_{(1)},b_{(2)}\ra \: x_{(2)} \tr f.
\eeq
Dropping the ``$\tr$'' we can write this for
arbitrary functions $f$ in the form of commutation
relations
\beq
\fbox{$x\;b = b_{(1)} \la x_{(1)},b_{(2)}\ra \: x_{(2)} .$}
\label{commrel}
\eeq
This commutation relation provides \x $\A \ot \U$\x
with an algebra structure
\beq
\begin{array}{l}
\cdot :\;(\A \ot \U) \ot (\A \ot \U) \rightarrow
\A \ot \U: \\
(a \ot x) \cdot (b \ot	y)
= a \: b_{(1)} \la x_{(1)},b_{(2)}\ra \ot x_{(2)} \: y.
\end{array}
\label{crossprod}
\eeq
The resulting associative algebra is the generalized semi-direct
product of \A\ and \U;
it is denoted \A \cross \U\ and we call it the {\em quantized algebra
of differential
operators.} The commutation relation (\ref{commrel})
should be interpreted as a product in $\A\cross\U$.
(From now on we will omit $\ot$-signs in \A\cross\U.)
This algebra contains all information about general vector fields,
their transformation properties and actions on functions.
It is bicovariant in the sense that it is a bi-$\A$-comodule
and a bi-$\U$-module (where the elements of $\U$ should be
interpreted as Lie derivatives.)

Equation (\ref{commrel}) can be used to calculate arbitrary inner products of
\U\ with \A\ , if we define a {\em right vacuum} ``$\ra $" to act like
the
counit in \U\ and a {\em left vacuum} ``$\la $" to act like the counit in \A
\beq \begin{array}{rcl}
\la x\: b\ra & = & \la b_{(1)} \la x_{(1)},b_{(2)}\ra \: x_{(2)}\ra \\
      & = &\epsilon (b_{(1)}) \la x_{(1)},b_{(2)}\ra \: \epsilon (x_{(2)}) \\
      & = & \la x,b\ra ,\z \mbox{for } \forall \x x \in \U ,\, b \in \A .
\end{array} \eeq
Using only the right vacuum we recover the formula for
left actions
\beq
\begin{array}{rcl}
x\: b \ra & = & b_{(1)} \la x_{(1)},b_{(2)}\ra	x_{(2)}\ra \\
      & = & b_{(1)} \la x_{(1)},b_{(2)}\ra  \epsilon (x_{(2)}) \\
      & = & b_{(1)} \la  x , b_{(2)}\ra \\
      & = & x \tr b,\z \mbox{for } \forall \x x \in \U ,\, b \in \A .
\end{array}
\label{rightvac}
\eeq
It can be shown that the right coaction of $\A$ on $\smash$
with the correct geometrical meaning is obtained through
{\em conjugation} by the canonical element $C$ of $\A\ot\U$
\beq
\DA(\alpha ) \equiv \alpha^{(1)} \ot \alpha^{(2)'}
	     = C(\alpha \ot 1)C^{-1} \label{CaC}
\eeq
for any $\alpha \in \smash$.  This expression shows explicitly that
$\DA$ is an algebra homomorphism
and that it is consistent with the algebra
structure of $\smash$. Given linear dual bases $\{ e_\alpha \}$ of \U\
and $\{ f^\alpha \}$ of \A\ the canonical element is $C = e_\alpha
\ot f^\alpha$.
We can also define a
left \U-coaction for elements of $\A\cross\U$
\beq
\UD(\alpha ) \equiv \alpha _{1'} \ot \alpha _{2}
= C^{-1} (1 \ot \alpha) C,
\eeq
that appears in the general commutation relation
\beq
\alpha \beta = \beta ^{(1)} \la \alpha _{1'},\beta ^{(2)'}\ra\alpha _{2}
\eeq
for arbitrary elements $\alpha, \beta \in \A\cross\U$.

\section{Cartan Calculus}

So far we have considered the dually paired Hopf algebras \U\ of
left-invariant operators (vector fields) and  \A\ of
functions on a quantum group.
{\em All} information about a Quantum Group is contained in $\smash$,
but to obtain more tools for applications and in particular to introduce
the concept of ``infinitesimality'', we still have to consider additional
structure.
We will extend the Hopf algebras $\U$ and $\A$
by formally adjoining spaces $\Ileft$ and $\Gleft$ respectively.
$\Ileft$ is spanned by a basis of
symbols $\{ \I_i \}$ of degree $-1$, called
{\em inner derivations}, similarly $\Gleft$
is spanned by a basis $\{ \om^i \}$
of left-invariant 1-forms with degree $+1$ and $< \I_i,\om^j > = \delta_i^j$.
In the following we shall derive consistency
conditions under which the spaces
\beq
\Omega = \A \ot \left( \bigoplus_{i = 1}^{\infty} \Gleft^{\ot i}
	 \right)
\eeq
and
\beq
{\cal D} = \U \ot \left( \bigoplus_{i = 1}^{\infty} \Ileft^{\ot i}
	   \right)
\eeq
are again (graded) Hopf algebras. The semi-direct product of these
Hopf algebras is an bicovariant
associative graded algebra of inner derivations,
Lie derivatives and forms, namely the Cartan Calculus\footnote{
We shall also require the existence of a biinvariant operator
$\dl$ of degree $+1$} \cite{Lin,AscCas,Ixt}.
This construction gives an economic way
to derive all commutation relations in the Cartan Calculus.
The possible commutation relations within $\Omega$ and $\cal D$ are for
instance restricted by aforementioned consistency conditions.

\subsection{Derivatives and Differential Forms}

\subsubsection{Graded Hopf Algebra of Derivatives}
\label{s:der}

Let us start by introducing an operation $\UD:\Ileft \rightarrow
\U \ot \Ileft$ on the space of inner derivations.
We would like to investigate under what conditions
\beq
\begin{array}{rcl}
\Delta(\I_i) & = &  \I_i \ot 1 + \UD(\I_i)\\
	     & =: &  \I_i \ot 1 + L_i{}^j \ot \I_j
\end{array}
\eeq
is a coproduct in $\cal D$ satisfying the Hopf algebra axioms:\\
$(\Delta \ot \id)\Delta = (\id \ot \Delta)\Delta$,
$\cdot(\ep \ot \id) = \id$, and $\cdot(\id \ot \ep) = \id$
require
\beq
\Delta L = L \dot{\otimes} L,\x \ep L = I,\x S L = L^{-1}, \x
\mbox{and }\ep(\I_i) = 0.
\eeq
$\cdot(S \ot \id) \Delta = \ep$ implies
\beq
S \I_i = - SL_i{}^j \I_j.	 \label{si}
\eeq
$\cdot(\id \ot S) \Delta= \ep$ is then automatically satisfied.
Next we will consider commutation relations between elements of \U\
and $\I$. As we have in mind to let the elements of \U\
act as Lie derivatives or Lie transports, the action of these
left-invariant differential operators on those
inner derivations should be given via a right \A-coaction
\beq
\Li_{x} \I_i = \I_j\la x_{(1)},T^j{}_i\ra \Li_{x_{(2)}} \label{lx}
\eeq
where $\DA(\I_i) = \I_j \ot T^j{}_i$ and $x \in
\U$.\footnote{``$\Li_{x}$'' reads: ``Lie derivative along the vector
field $x$''.}

\noindent {\em Remark:} The form of (\ref{lx}) can also
be derived as follows: Introduce a right \U-coaction $\DU$, such that
$\DU \equiv \Delta$ on \U\ and $\DU(\I_i) = \I_i \otimes 1$, then
make an ansatz for the $\Li$--$\I$ commutation relations that
is linear and homogeneous in $\I$:
$\Li_x \I_i = \I_j \Li_{\tilde{T}^j{}_i(x)},$ where
$\tilde{T}^j{}_i:\U \rightarrow \U$ are	linear maps.
Requiring covariance under $\DU$ gives
$\tilde{T}^j{}_i(x) =	\ep(\tilde{T}^j{}_i(x_{(1)}) ) x_{(2)}$
or $\tilde{T}^j{}_i(x) = \la x_{(1)}, T^j{}_i \ra x_{(2)}$ with
$T^j{}_i := f^{\al} \ep( \tilde{T}^j{}_i (e_{\al}) ) \in \A$.

The requirement that $\DA$ be a coaction and in particular an
algebra homomorphism immediately gives $\Delta T = T \dot{\otimes} T$,
$\ep T = I$, and $S T = T^{-1}$. It also put constraints on the $T$--$T$
commutation relations. We will come back to that (\ref{at}).
The coproduct $\Delta$ must be a homomorphism in the extended algebra
$\cal D$, \ie it should preserve all its
commutation relations. The $\Li$--$\I$ commutation relations
give
\beq
\Delta(\Li_{x}) \UD(\I_i)
= \UD(\I_j) \la x_{(1)}, T^j{}_i\ra \Delta(\Li{x_{(2)}}),
\eeq
\ie that $\UD$ must be
a coaction and in particular a homomorphism of the
$\Li$--$\I$ commutation relations.
For the specific form of the coaction on the $\I$'s we find:
\beq
x_{(1)} L_i{}^j \la x_{(2)},T^k{}_i\ra =
\la x_{(1)},T^j{}_i\ra L_j{}^k x_{(2)}	 \label{xl}
\eeq
or, with $x = L$ and $\hat{R}^{ab}{}_{cd} := \la L_c{}^b,T^a{}_d\ra $,
\beq
\hat{R}_{12} L_1 L_2 = L_2 L_1 \hat{R}_{12}.  \label{rll}
\eeq
The commutation relations between inner derivations should be
homogeneous to preserve the grading. They can therefore be written
in terms of projection operators on quadratic and perhaps higher
order\footnote{$GL_q(n)$, $SL_q(n)$ and $SU_q(n)$ only have quadratic
relations. For $SO_q(n)$ see \cite{SOn}.}
combinations of inner derivations. Here we would like to consider
quadratic relations with projection operators $P^{(\al)}$.
\beq
\Delta(\I_i \I_k) = \I_i \I_k \otimes 1 + \UD(\I_i) (\I_k \otimes 1)
+ (\I_i \otimes 1) \UD(\I_k) + \UD(\I_i) \UD(\I_k)
\label{dii}
\eeq
must be consistent with relations of the form
\beq
\I_i \I_k P^{(\al)}{}^{ik}{}_{st} = 0.
\eeq
The first term of (\ref{dii}) is trivially consistent. The fourth
term requires $\UD$ to be a homomorphism of the $\I$--$\I$ relations
and hence (recalling our previous results and the fact that
$\Delta \equiv \UD$ on \U,) a homomorphism of all of $\cal D$.
We will see that this is indeed satisfied after considering the
second and third terms in (\ref{dii}):
\beq
\bar
0 & = & \left(\UD(\I_i) (\I_k \otimes 1)
      + (\I_i \otimes 1) \UD(\I_k)\right) P^{(\al)}{}^{ik}{}_{st}\\
  & = & \left(-L_i{}^j \I_k \otimes \I_j + \I_i L_k{}^l \otimes \I_l\right)
      P^{(\al)}{}^{ik}{}_{st}\\
  & = & \I_r L_j{}^l \otimes \I_l
      (\delta_i^r \delta_k^j - \la L_i{}^j,T^r{}_k \ra)
      P^{(\al)}{}^{ik}{}_{st}\\
  & = & \I_r L_j{}^l \otimes \I_l
      (I - \hat{R})^{rj}{}_{ik}
      P^{(\al)}{}^{ik}{}_{st},			  \label{iip}
\ear
\eeq
\ie
\beq
(I - \hat{R}) P^{(\al)} = 0 \mbox{ for all $\al$}. \label{rp}
\eeq
Conditions on the projectors of higher order relations of the $\I$'s
are obtained analogously. However, do to the characteristic equation
for $\hat{R}$ these relations may already
be implied by the quadratic relations.
The ``$-$''-sign in the second line of (\ref{iip}) comes from
the graded tensor product structure of $\cal D$:
\beq
(a \otimes b)(c \otimes d) = (-1)^{{\rm deg}(b)\cdot {\rm deg}(c)}
 a c \otimes b d.
\eeq
Note that {\em all} $P^{(\al)}$ have to satisfy (\ref{rp}). Adding
additional relations ``by hand'', as it was done in \cite{SOn},
breaks the structure under consideration and also leads to problems
for higher order relations \cite{IsPy}.
The homomorphism of $\UD$ with respect to the last term in (\ref{dii})
can be seen in two ways: In view of equation (\ref{rp}) we can represent
$\I_i \I_j$ as $\I_k \otimes \I_l (I - \hat{R})^{kl}{}_{ij}$ and then
we get $L_m{}^i L_n{}^j \otimes \I_i \I_j P^{(\al)}{}^{mn}{}_{op}$ =
$L_m{}^i L_n{}^j \otimes (\I_k \otimes \I_l (I - \hat{R})^{kl}{}_{ij})
P^{(\al)}{}^{mn}{}_{op}$ = $L_i{}^k L_j{}^l \otimes (\I_k \otimes \I_l)
(I - \hat{R})^{ij}{}_{mn} P^{(\al)}{}^{mn}{}_{op}$ = $0$, where we have
used equation (\ref{rll}). Alternatively we could write $P^{(\al)}$
in terms of $\hat{R}$ employing its characteristic equation; equation
(\ref{rll}) will then give the desired result again.

All that is left to check for the Hopf algebra structure of
$\cal D$ is the antipode and the counit. The counit of \U\ extends
trivially to a homomorphism of all of $\cal D$ because of
$\ep(\I_i) = 0$. The antipode should be a graded anti-homomorphism
of $\cal D$:
\beq
S(a b) = (-1)^{{\rm deg}(a)\cdot {\rm deg}(b)} S(b) S(a).
\eeq
Consistency of the $\Li$--$\I$ commutation relation:
\beq
\bar
S(\Li_x \I_i) & = & S( \I_l \la x_{(1)}, T^l{}_i \ra \Li_{x(2)} )\\
& = & -S(\Li_{x_{(2)}}) \la x_{(1)} , T^l{}_i \ra SL_l{}^k \I_k\\
& = & -SL_i{}^l S(\Li_{x_{(1)}}) \la x_{(2)}, T^k{}_l \ra \I_k\\
& = & -SL_i{}^l \la x_{(3)}, T^k{}_l \ra \I_n
      \la S x_{(2)}, T^n{}_k \ra S(\Li_{x_{(1)}})\\
& = & -SL_i{}^l \I_l S(\Li_x) \: = \: S(\I_i) S(\Li_x),
\ear
\eeq
where equation (\ref{xl}) was used in the second line.
The consistency of $\I$--$\I$ type relations can be reduced
to the requirement that $\UD$ be a homomorphism:
\beq
\bar
S(\I_i \I_j) P^{(\al)}{}^{ij}{}_{op} & = & -SL_j{}^k \I_k SL_i{}^l
    \I_l P^{(\al)}{}^{ij}{}_{op}\\
& = & -SL_j{}^k SL_i{}^n \I_m \I_l \hat{R}^{ml}_{nk}
    P^{(\al)}{}^{ij}{}_{op}\\
& = & -SL_k{}^l SL_n{}^m \I_m \I_l \hat{R}^{nk}_{ij}
    P^{(\al)}{}^{ij}{}_{op}\\
& = & -S(L_n{}^m L_k{}^l) \I_m \I_l P^{(\al)}{}^{nk}{}_{op} \: = \: 0.
\ear
\eeq
Given an $L \in M_n(\U)$, $n = $dim$(\Ileft)$
that forms a representation of \A\ and satisfies equation
(\ref{xl}) and a $T \in M_n(\A)$ (see below)
we have hence succeeded in extending the
Hopf algebra structure of \U\
to $\cal D$ while obtaining natural commutation relations
among the elements of $\Ileft$ on the way.

\subsubsection{Graded Hopf Algebra of Differential Forms}

In pretty much the same way as for \U\ we can extend the Hopf algebra
structure of \A\ to $\Omega$ if we introduce a $\DA$ and a $\Delta$
on $\Gleft$:
\begin{eqnarray}
\DA(\om^i) & = & \om^j \otimes A^i{}_j,\label{dom}\\
\Delta(\om^i) & = & \DA(\om^i) + 1 \otimes \om^i,\\
S(\om^i) & = & -\om^j SA^i{}_j,\\
\ep(\om^i) & = & 0.
\end{eqnarray}
The multiplication between elements of \A\ and $\Gleft$
should be covariant under a left \A-coaction $\AD$ and is hence
given via a right \U-coaction $\DU(\om^i) = \om^j \otimes B_j{}^i$
(see the remark in section~\ref{s:der}):
\beq
a \om^i = \om^j B_j{}^i(a).\label{omba}
\eeq
(In \cite{SrZ} equation (\ref{dom}) is analyzed from the
point of view of inhomogeneous quantum groups and (\ref{omba})
is interpreted as a braiding arising from an universal $\cal R$
in the quantum double of $\U$.)
Having in mind to combine $\Omega$ and $\cal D$ into a semi-direct
product algebra $\Omega\cross {\cal D}$ we would like them to
be dually paired Hopf algebras with respect to an inner
product $<\:,\:>: {\cal D} \otimes \Omega \rightarrow k$
in the sense of
\begin{eqnarray}
< x y, a > & = & (-1)^{{\rm deg}(x) \cdot {\rm deg}(y)}
 < x, a_{(1)} > < y,a_{(2)}>,\\
< x, a b> & = & (-1)^{{\rm deg}(a) \cdot {\rm deg}(b)}
 < x_{(1)},a > < x_{(2)},b >,\\
< S x, a > & = & < x, S a >.
\end{eqnarray}
On $\U \otimes \A$ this new inner product reduces to the
old one ($\la\:,\:\ra$) with $\Li_x$ interpreted as $x \in \U$,
furthermore it is zero unless
the overall degree of the elements in $<\:,\:>$ is zero.
The only thing left to fix is: $<\I_i,\om^j> = \delta_i^j$.

$A$ and $B$ are no longer independent from $L$ and $T$:
\beq
\bar
\ep(x) \delta^i_j & = &
   <\Li_x \I_k, \om^i>\\
   & = & <\I_l \la x_{(1)},T^l{}_k\ra \Li_{x_{(2)}}, \om^i >\\
   & = & \delta_l^j \la x_{(1)},T^l{}_k \ra \la x_{(2)},A^i{}_j\ra\\
   & = & \la x, T^j{}_k A^i{}_j \ra,
\ear
\eeq
\ie $A = S^{-1} T$.
\beq
\bar
\la L_k{}^i,a \ra & = & <\I_k, a \om^i>\\
   & = & <\I_k, \om^j B_j{}^i(a)>\\
   & = & \la B_k{}^i,a \ra,
\ear
\eeq
\ie $B = L$.

It has been known for some time \cite{Sud} that the Woronowicz type calculus
of forms and functions
on $A_n$ quantum groups forms a Hopf algebra. Here we are just
reversing the logic: We {\em require} the calculus to be a graded
Hopf algebra and then derive consistency relations from this.
Following is a summary of definitions and relations that are needed
to turn both $\cal D$ and $\Omega$ into Hopf algebras:

\paragraph{Dually Paired $\cal D$- and $\Omega$-Hopf Algebras (Summary)}

\begin{eqnarray}
\UD(\I_i) & = & L_i{}^j \otimes \I_j\\
\DU(\om^i) & = & \om^j \otimes L_j{}^i\\
\DA(\I_i) & = & \I_j \otimes T^j{}_i\\
\DA(\om^i) & = & \om^j \otimes S^{-1}T^i{}_j\\
\Delta(L) & = & L \dot{\otimes} L\\
\ep(L) & = & I \\
S(L) & = & L^{-1}\\
\Delta(T) & = & T \dot{\otimes} T\\
\ep(T) & = & I \\
S(T) & = & T^{-1}\\
x_{(1)} L_i{}^j \la x_{(2)},T^k{}_j\ra & = &
     \la x_{(1)},T^j{}_i\ra L_j{}^k x_{(2)}\\
a_{(1)} T^k{}_j \la L_i{}^j, a_{(2)} \ra
     & = & \la L_l{}^k, a_{(1)} \ra T^l{}_i a_{(2)} \label{at}\\
\hat{R}^{ij}{}_{kl} & := & \la L_k{}^j, T^i{}_l \ra \\
\hat{\sigma}^{ij}{}_{kl} & := & \la L_k{}^j, S^{-1}T^i{}_l \ra
     \: = \: \hat{R}^{-1}\\
\I_i \I_j P^{(\al)}{}^{ij}{}_{kl} = 0 %, (P^{(\al)})^2 = P^{(\al)}
     & \Rightarrow & (I - \hat{R}) P^{(\al)} = 0\\
P_{(\al)}{}^{ij}{}_{kl} \om^k \om^l = 0 %, (P_{(\al)})^2 = P_{(\al)}
     & \Rightarrow & P_{(\al)} (I - \hat{\sigma}) = 0
\end{eqnarray}
Equations similar two the last two relations are also true for
higher order relations in the $\I$'s and $\om$'s.

\subsection{Semidirect Product of $\Omega$ and ${\cal D}$}

Having obtained dually paired graded Hopf algebras $\Omega$
and ${\cal D}$ we will now construct a bi-\A-cocovariant
(and bi-\U-covariant) algebra of Lie derivatives, inner derivations,
functions, and forms from their semi-direct product. The multiplication
in $\Omega\cross{\cal D}$ is given by
\beq
x a = (-1)^{{\rm deg}(x) \cdot {\rm deg}(a)}
a_{(1)} (a_{(2)},x_{(1)}) x_{(2)}
\eeq
where $x \in {\cal D}$, $a \in \Omega$ and
\beq
(b,y) \equiv (-1)^{{\rm deg}(b)\cdot{\rm deg}(y)} <y,b>,\z{\rm for all }
y \in {\cal D},~b \in \Omega.
\eeq
In the calculation of
the grading that we  use one acquires a minus sign whenever
an odd quantity moves past another odd quantity and
$<y,b> = 0$ unless ${\rm deg}(y)
+ {\rm deg}(b) = 0$. (A graphical presentation of the graded
semi-direct product or rather its braided generalization is
given below.) The grading was chosen such that
$\Omega\cross{\cal D}$ is an associative algebra.
Actions of ${\cal D}$ on $\Omega$ can be recovered either
via the adjoint action
\beq
x(a) \equiv x \tr a = (-1)^{{\rm deg}(a) \cdot {\rm deg}(x_{(2)})}
 x_{(1)} a S(x_{(2)})
\eeq
or through a right \U-vacuum that acts like the counit on $\cal D$:
\beq
\bar
x a > & = & (-1)^{{\rm deg}(x) \cdot {\rm deg}(a)}
a_{(1)} (a_{(2)},x_{(1)}) x_{(2)} >\\
 & = & (-1)^{{\rm deg}(x) \cdot {\rm deg}(a)}
a_{(1)} (a_{(2)},x_{(1)}) \ep(x_{(2)})\\
 & = & (-1)^{({\rm deg}(a) + {\rm deg}(x)) \cdot {\rm deg}(x)}
a_{(1)} <x,a_{(2)}> .
\ear
\eeq
Inner products can be calculated from the commutation relations
with the additional help of a left \A-vacuum that acts like the counit on
$\Omega$:
\beq
\bar
< x a > & = & (-1)^{{\rm deg}(x) \cdot {\rm deg}(a)}
 < a_{(1)} (a_{(2)},x_{(1)}) x_{(2)} >\\
 & = & (-1)^{{\rm deg}(x) \cdot {\rm deg}(a)}
 \ep(a_{(1)}) (a_{(2)},x_{(1)}) \ep(x_{(2)})\\
 & = &  +<x,a> .
\ear
\eeq
The proof of the bicovariance of the semidirect product algebra
is virtually identical to the ungraded case \cite{} and therefore
left as an exercise.
Following is the graphical picture of the semi-direct product
construction and
an example of the computation of a commutation relation:\\[2ex]
\unitlength=1.00mm
%\special{em:linewidth 0.4pt}
\linethickness{0.4pt}
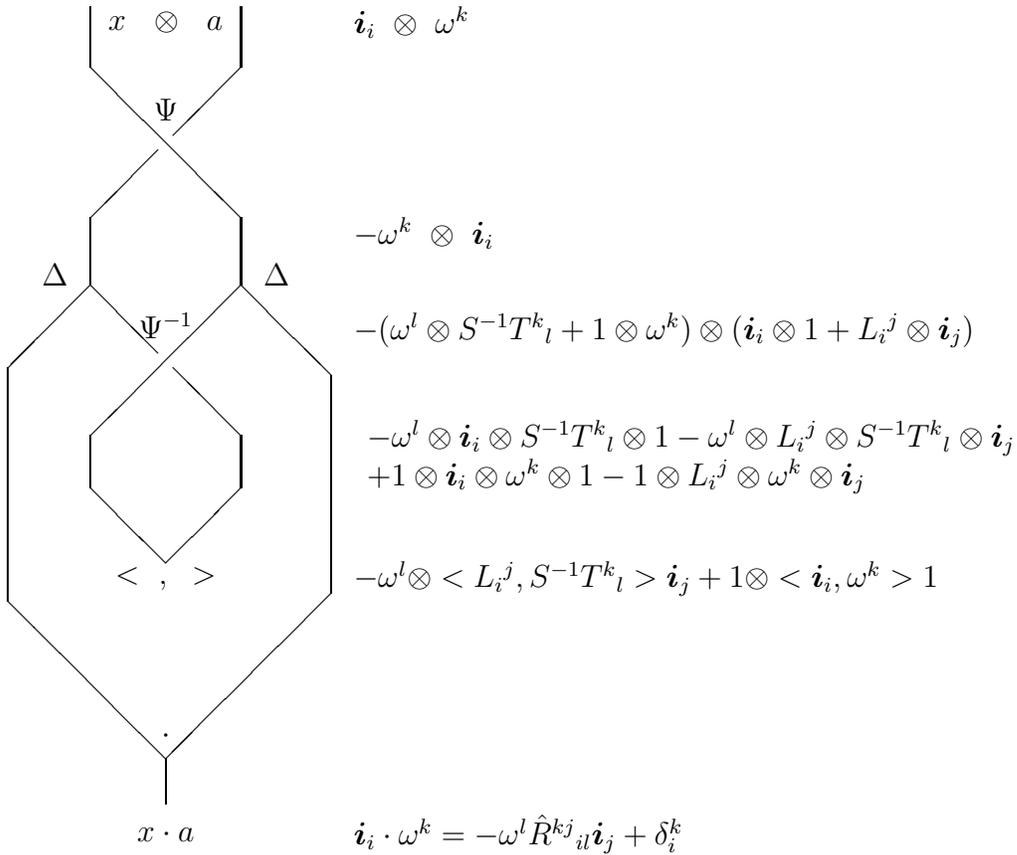
\begin{picture}(47.00,112.00)
\put(12.00,104.00){\line(1,-1){20.00}}
\put(32.00,84.00){\line(0,-1){9.00}}
\put(32.00,75.00){\line(1,-1){12.00}}
\put(44.00,63.00){\line(0,-1){29.00}}
\put(44.00,34.00){\line(-1,-1){22.00}}
\put(32.00,104.00){\line(-1,-1){9.00}}
\put(21.00,93.00){\line(-1,-1){9.00}}
\put(12.00,84.00){\line(0,-1){9.00}}
\put(12.00,75.00){\line(-1,-1){11.00}}
\put(22.00,12.00){\line(-1,1){21.00}}
\put(1.00,33.00){\line(0,1){31.00}}
\put(32.00,75.00){\line(-1,-1){20.00}}
\put(32.00,55.00){\line(-1,1){9.00}}
\put(12.00,75.00){\line(1,-1){9.00}}
\put(12.00,55.00){\line(0,-1){7.00}}
\put(12.00,48.00){\line(1,-1){10.00}}
\put(22.00,38.00){\line(1,1){10.00}}
\put(32.00,48.00){\line(0,1){7.00}}
\put(35.00,75.00){\makebox(0,0)[lb]{$\Delta$}}
\put(9.00,75.00){\makebox(0,0)[rb]{$\Delta$}}
\put(22.00,97.00){\makebox(0,0)[cb]{$\Psi$}}
\put(22.00,68.00){\makebox(0,0)[cb]{$\Psi^{-1}$}}
\put(22.00,36.00){\makebox(0,0)[cc]{$<~,~>$}}
\put(22.00,15.00){\makebox(0,0)[cc]{$\cdot$}}
\put(47.00,110.00){\makebox(0,0)[lc]{$\I_i ~\otimes ~\om^k$}}
\put(47.00,84.00){\makebox(0,0)[lt]{$-\om^k ~\otimes ~\I_i$}}
\put(47.00,69.00){\makebox(0,0)[lc]{$-(\om^l \otimes S^{-1}T^k{}_l
+ 1 \otimes \om^k) \otimes (\I_i \otimes 1 + L_i{}^j \otimes \I_j)$}}
\put(47.00,52.00){\makebox(0,0)[lc]{$\begin{array}{l}
-\om^l\otimes\I_i\otimes S^{-1}T^k{}_l\otimes 1
-\om^l\otimes L_i{}^j\otimes S^{-1}T^k{}_l\otimes \I_j\\
+1\otimes \I_i \otimes \om^k \otimes 1 -1\otimes L_i{}^j
\otimes \om^k\otimes\I_j\end{array}$}}
\put(47.00,36.00){\makebox(0,0)[lc]{$-\om^l \otimes
<L_i{}^j,S^{-1}T^k{}_l> \I_j + 1\otimes <\I_i,\om^k>1$}}
\put(47.00,2.00){\makebox(0,0)[lc]{$\I_i \cdot \om^k =
-\om^l \hat{R}^{kj}{}_{il} \I_j + \delta^k_i$}}
\put(32.00,104.00){\line(0,1){8.00}}
\put(12.00,112.00){\line(0,-1){8.00}}
\put(22.00,110.00){\makebox(0,0)[cc]{$x~~\otimes~~a$}}
\put(22.00,2.00){\makebox(0,0)[cc]{$x\cdot a$}}
\put(22.00,12.00){\line(0,-1){6.00}}
\end{picture}
\\
(The line crossings indicate how to generalize to the
case where $\Psi$ describes a true {\em braiding}\/ rather than
a grading.)

We can see that the term $1 \otimes \om^k$ in $\Delta(\om^k)$
is necessary for $\I_i(\om^k) = \delta_i^k$. We will later learn
about the ``meaning'' of the other term ($\DA(\om^k)$).

\subsection{Some Remarks on the Exterior Derivative}

So far we have constructed a consistent bicovariant algebra
of derivatives and forms. {\em This}\/ is the rigid framework
for a differential geometry of quantum groups.
It is now of interest to investigate
the possibility of a biinvariant map $\dl:\A \rightarrow \Gleft$,
called the exterior derivative (on functions).
Biinvariance means
\beq
\AD \circ \dl = (\id \otimes \dl) \Delta,\x
\DA \circ \dl = (\dl \otimes \id) \Delta.
\eeq
$\dl(a)$ should be of first order in the $\om^i$ and linear in $a$, \ie
$\dl(a) = \om^i \tilde{\chi}_i(a)$ with  linear maps $\tilde{\chi}_i:
\A \rightarrow \A$. The requirement of left-invariance leads to
\beq
\tilde{\chi}_i(a)  =  a_{(1)} \ep(\tilde{\chi}_i(a_{(2)}))
   =  a_{(1)} \la \chi_i, a_{(2)} \ra
\eeq
where $\chi_i = e_\al \tilde{\chi}_i(f^\al )$. (We will drop the
``$\tilde{\x}$'' on $\tilde{\chi}_i$ from here on.)
We can learn a couple of things from this equation:
\newcounter{things}
\begin{list}{\roman{things})}{\usecounter{things}
\setlength{\rightmargin}{\leftmargin}}
\item $\dl$ satisfies
the Leibniz rule on $\A$  iff $\Delta\chi_i =
\chi_i \otimes 1 + L_i{}^j \otimes \chi_j.$
\item $\I_i(\dl(a)) = \I_i(\om^j \chi_j(a)) = \chi_i(a)$, \ie
$\I_i \equiv \I_{\chi_i}$.
\item $\dl$ can be written as an operator $\om^i \chi_i$ on \A.
\end{list}
We also find \cite{Sud}
\beq
\Delta(\dl a)  = \dl(a_{(1)}) \otimes a_{(2)} + a_{(1)} \otimes
\dl(a_{(2)}) \label{dom2}
\eeq
essentially because $\dl$ is biinvariant.
The first term of this equation implies
\beq
\Li_x \circ \dl = \dl \circ \Li_x
\eeq
--- and in particular $L_i{}^j \circ \dl = \dl \circ L_i{}^j$\footnote{That
is why we chose the letter ``$L$'' in the first place.} ---,
the second term implies
\beq
\Lix{i}(a) = \I_i(\dl a).
\eeq

\subsection{Explicit Realizations of all this}

Right invariance of $\dl$ imposes another condition on $\chi_i$:
\beq
\DA(\chi_i) = \chi_j \otimes T^i{}_j,
\eeq
where $\DA(\chi_i) \equiv e_{\al(1)} \chi_i S(e_{\al(2)}) \otimes
f^\al$ from (\ref{CaC}).
All together there are quite a few conditions on the objects
under consideration and one may wonder whether they can all be
satisfied.
Luckily that is so and there are even (at least) two possibilities:

\paragraph{Extrinsic Braiding --- Plane Type Calculi}

For quasitriangular Hopf algebras one may chose
$L_j{}^i$ proportional to either $\la \R,T^i{}_j \otimes \id \ra$
or $\la \R, \id \otimes S^{-1} T^i{}_j \ra$, where
$T$ is the adjoint representation corresponding to a set
of bicovariant generators $\{ \chi_i \}$.
The corresponding $f$--$\om$ relations are used in \cite{PyFa}
to reduce the number of 1-forms. That is possible because
$L$ and $T$ are simultaneously reducible in this formulation.
The price one has to pay is that the Leibniz rule for
$\dl$ will no longer be satisfied.

The commutation relations of quantum planes are incidentally derived
from an $L$ that is of the above form \cite{WZ}.

\paragraph{Intrinsic Braiding --- Woronowicz \protect\cite{Wor}
Type Calculi}

Here one assumes again the existence of a set
of bicovariant generators $\{ \chi_i \}$ but further requires
a coproduct of the form
$\Delta\chi_i =\chi_i \otimes 1 + L_i{}^j \otimes \chi_j,$
so that the Leibniz rule for $\dl$ is satisfied.
Given Hopf algebras \A\ and \U\ it is always possible to
find such generators. The trivial choice being the
elements of a (formal) linear basis $\{ e_\al \}$ of \U\
--- except for the element ``1'' of course. This choice
will then lead to a universal Cartan calculus \cite{SW}.
This universal Calculus can be reduced in dimensionality
by dividing by certain ideals. One
typically ends up with more generators than in the classical
case though.
In the following we will assume that we are dealing with
this type of calculus. An explicit realization for quantum
groups of the type described in \cite{FRT} was given
by B.\ Jurco \cite{Jur}.

Now we have all the tools necessary to investigate the meaning
of $S(\I_i)$:

\subsection{Antipode of Inner Derivations}

When we calculate the {\em action} of $S(\I_i)$ on functions and
differentials we find
\beq
S\I_i(f) = 0,\x S\I_i(\dl f) = S\chi_i(f),
\eeq
as expected. If we however calculate commutation relations
using $S\I_i = -SL_i{}^j\I_j$, we are in for a surprise:
\beq
S\I_i f = f S\I_i,
\eeq
\ie $S\I$ commutes with functions --- there is no braiding {\em here} ---,
but
\beq
S\I_i \dl(f) = - \dl(f) S\I_i + S\chi_j(f) SL_i{}^j,\label{but}
\eeq
so $S\I$ braids as it acts on something, not when it moves through
a term.
This is the {\em post}-grading calculus as opposed to the
standard {\em pre}-grading calculus of the $\I$'s.
In effect (\ref{but}) should be read backwards, i.e. from right
to left.
To find an interpretation of $S\I$ we have to write the equations
in a more standard way. As all what really matters are inner
products (``matrix elements'') we can achieve this goal
by introducing a {\em left}\/ \U-vacuum and a {\em right}
\A-vacuum together with a left-acting $\stackrel{\leftarrow}{S\I_i}$
and a new commutation relation
\beq
\dl(f) \stackrel{\leftarrow}{S\I_i} = S\chi_i(f) -
\stackrel{\leftarrow}{S\I_j} SL_i{}^j(\dl f).
\eeq
A real nice feature is that $\stackrel{\leftarrow}{S\I_i}$ is already
realized in $\Omega\cross {\cal D}$:
Multiply equation (\ref{but}) by $S^2L_j{}^i$ from the right to find
after some manipulations using $S^2\I_i = S\I_j S^2 L_i{}^j$ that
\beq
\dl(f) (-S^2\I_i) = S\chi_i(f) - (-S^2\I_j) SL_j{}^i(\dl f),
\eeq
\ie
\beq
\stackrel{\leftarrow}{S\I_i} \equiv -S^2\I_i.
\eeq
The way it acts, $S\I$ should be interpreted as a right-invariant
object. From its realization in $\Omega\cross {\cal D}$ we
can compute it's left transformation property:
\beq
\AD(S\I_i) = S^{-2}T^j{}_i \otimes S\I_j.
\eeq

\noindent {\em Remark:} This is also a solution to the problem
that the antipodes of a set of $\{ \chi_i \}$ that closes
under right coactions does not in general close under $\DA$
again. If interpreted as right-invariant and left-acting
objects, the $\{ S\chi_i \}$ {\em will} however close
under $\AD$.

\subsection{Cartan Identity}

The graded Hopf algebra of $\Omega$ can be very easily expressed
if one treats $\dl$ as an independent element with
\begin{eqnarray}
\Delta(\dl) & = & \dl \otimes 1 + 1 \otimes \dl,\\
S(\dl) & = & -\dl,\\
\ep(\dl) & = & 0.
\end{eqnarray}
The exterior derivative $\dl$ can now be viewed as an additional
operator in $\cal D$; equations (\ref{dom},\ref{dom2})
follow then automatically.
We would now like to proof the following interesting relation
among Lie derivatives, inner derivations, and the exterior
derivative
\beq
\Li_{\chi_k} = \dl \I_k + \I_k \dl,\label{cartid}
\eeq
due to Cartan.
We start by checking the coproduct:
\beq
\bar
\Delta(\dl \I_k + \I_k \dl) & = & (\I_k \otimes 1
+ L_k{}^l \otimes \I_l)(\dl \otimes 1 + 1 \otimes \dl)\\
&&+(\dl \otimes 1 + 1 \otimes \dl)(\I_k \otimes 1
+ L_k{}^l \otimes \I_l)\\
 & = & (\dl \I_k + \I_k \dl)	\otimes 1 + L_k{}^l \otimes
(\dl \I_l + \I_l \dl). \label{coplie}
\ear
\eeq
Similar one checks $S$ and $\ep$.
Using relations that we have previously derived we
see that
\beq
(\dl \I_i + \I_i \dl)(f) = 0 + \I_i(\dl f) = \chi_i(f)
\eeq
and
\beq
(\dl \I_i + \I_i \dl)(\dl f) = \dl(\chi_i(f))
\eeq
thus verifying (\ref{cartid}) on $f$ and $\dl f$.
Using the coproduct (\ref{coplie}) in $\cal D$ this immediately extends to
all of $\Omega$.

\subsection{Summary of Relations in the Cartan Calculus \protect\cite{Ixt}}

\paragraph{Commutation Relations}

For any $p$-form $\alpha $:
\begin{eqnarray}
\dl \alpha   & = & \dl(\alpha ) + (-1)^{p} \alpha  \dl\\
\I_{\chi _{i}} \alpha  & = & \I_{\chi _{i}}(\alpha ) + (-1)^{p}
L_{i}{}^{j}(\alpha ) \I_{\chi _{j}}\\
\Li_{\chi _{i}} \alpha  & = & \Li_{\chi _{i}}(\alpha ) +
L_{i}{}^{j}(\alpha ) \Li_{\chi _{j}} \label{il1}
\end{eqnarray}

\paragraph{Actions}

For any function $f \in \A$, 1-form $\omega _{f} \equiv Sf_{(1)} \dl f_{(2)}$
and
vector field $\phi  \in \A\cross\U$:
\begin{eqnarray}
\Ix{i}(f) & = & 0\\
\Ix{i}(\dl f) & = & \dl f_{(1)}<\chi _{i},f_{(2)}>\\
\Ix{i}(\omega _{f}) & = & -<\chi _{i},S f>\\
\Lix{{}}(f) & = & \chi (f) = f_{(1)}<\chi ,f_{(2)}>\\
\Lix{{}}(\omega _{f})& = & \omega _{f_{(2)}} <\chi ,S(f_{(1)}) f_{(3)}>\\
\Lix{{}}(\phi )& = & \chi _{(1)} \phi  S(\chi _{(2)}) \label{il2}
\end{eqnarray}

\paragraph{Graded Quantum Lie Algebra of the Cartan Generators}

\begin{eqnarray}
\dl \dl & = & 0\\
\dl \Lix{{}} & = & \Lix{{}} \dl\\
\Li_{\chi _{i}} & = & \dl \I_{\chi _{i}} + \I_{\chi _{i}} \dl\\
\left[\Lix{i},\Lix{k} \right]_{q} & = & \Lix{l} f_{i}{}^{l}{}_{k}\\
\left[\Lix{i},\Ix{k} \right]_{q} & = & \Ix{l} f_{i}{}^{l}{}_{k}
\end{eqnarray}
Where $f_i{}^l{}_k = \la \chi_i , T^l{}_k \ra$
and $[\, , \,]_{q}$ is defined as follows
\begin{equation}
\left[\Lix{i}, \Box \right]_{q} :=
\Lix{i} \Box - L_i{}^j(\Box) \Lix{j}.
\end{equation}
This quantum Lie algebra becomes infinite dimensional as soon as we
introduce derivatives along general vector fields (see below).

\subsection{Lie Derivatives Along General Vector Fields}

We have seen in the introduction that general vector fields are
of the form $a^i \chi_i$, where $a^i \in \A$ are functional
coefficients. Inner derivations contract forms with vectors
``at each point''. Classically the $a^i$ are just numerical
coefficients at each point of the group manifold.
Here this property is expressible as
\beq
\I_{a^i \chi_i}  =  a^i \I_{\chi_i}.
\eeq
Through the Cartan identity
\beq
\Li_{a^i \chi_i}  =  \dl \I_{a^i \chi_i} + \I_{a^i \chi_i} \dl
\eeq
we can also consistently\footnote{On all forms.} introduce
Lie derivatives along general vector fields.
Combining these two equations we find
\beq
\Li_{a^i \chi_i} = a^i \Li_{\chi_i} + \dl(a^i) \I_{\chi_i}
\eeq
as in the classical case.

\subsection{A glimpse at extended calculi on the plane}

In the case of GL${}_q(n)$ the (bicovariant) vector fields
$\chi_i$ find a simple realization in the differential
operators of the linear quantum plane that can be used
to induce a Cartan calculus on the plane from the one of the
group \cite{CSZ}.
This extended calculus on the plane has some unexpected features like
for instance the appearance of differentials and inner derivations
in the commutation relations of Lie derivatives with functions.
For other quantum groups the $\chi_i$'s have non-linear
realizations on the plane and an extended calculus has not yet been
successfully induced.

It would be nice if the formalism presented in this lecture
could be generalized to the case of quantum planes
(\ie pseudo-Hopf algebras with non-trivial braiding)
to give an independent approach to the construction of extended
calculi.

\section*{Acknowledgments}

The geometric considerations in the introduction stem from
conversations with Chryss Chryssomalakos and Bruno Zumino  at the
ITP at Santa Barbara in the spring of 1993. The relations of
the Cartan calculus were developed throughout 1992
(in a different way) \cite{Lin,Ixt} (see also \cite{AscCas}) in collaboration
with Paul Watts and Bruno Zumino, who I would also like to thank
for many beneficial discussions about the subject of {\em this} lecture.
Last but not least I would like to thank Anthony Sudbery for
hospitality in York and many conversations that contributed much to
this work.

\end{document}